\documentclass[intlimits,twoside,a4paper]{article}
\usepackage{amsmath,amssymb}
\usepackage{graphicx}
\usepackage{wrapfig}
\usepackage[T2A]{fontenc}
\usepackage[cp1251]{inputenc}

\usepackage{cmpj2}

\def\be{\begin{equation}}
\def\ee{\end{equation}}
\def\bea{\begin{eqnarray}}
\def\eea{\end{eqnarray}}

\def\gsim{\mathrel{\lower.65ex\hbox{$\mathop{\kern0pt\sim}\limits
   ^{\lower.55ex\hbox{$>$}}$}}}
\def\lsim{\mathrel{\lower.65ex\hbox{$\mathop{\kern0pt\sim}\limits
   ^{\lower.55ex\hbox{$<$}}$}}}

\issue{2015}{18}{1}{13604}
\doinumber{10.5488/CMP.18.13604}

\title[Phase behaviour of DLVO model for lysozyme
and $\gamma$-crystalline aqueous electrolyte solutions]
{Theoretical study of phase behaviour of DLVO model for lysozyme
and $\gamma$-crystalline aqueous electrolyte solutions}

\author[R. Melnyk]{R. Melnyk}

\address{
Institute for Condensed Matter Physics
of the National Academy of Sciences of Ukraine,\\
1 Svientsitskii St., 79011 Lviv, Ukraine }
\date{Received January 12, 2015, in final form March 6, 2015}%

\begin{document}

\maketitle

\begin{abstract}
Mean spherical approximation (MSA), second-order Barker-Henderson (BH) perturbation theory
and thermodynamic perturbation theory (TPT) for associating fluids in combination with
BH perturbation theory are applied to the study of
the structural properties and phase behaviour of the Derjaguin-Landau-Verwey-Overbeek (DLVO) model of lysozyme and
$\gamma$-cristalline aqueous electrolyte solutions. Predictions of the MSA for the
structure factors are in good agreement with the corresponding computer simulation
predictions. The agreement between theoretical results for the liquid--gas phase diagram
and the corresponding results of the experiment and computer simulation is less satisfactory,
with predictions of the combined BH-TPT approach being the most accurate.

\keywords DLVO model, Yukawa potential, lysozyme
and $\gamma$-crystalline aqueous electrolyte solutions,
mean-spherical approximation
\pacs 64.10.+h, 61.46.-w, 05.70.Fh
\end{abstract}

\section{Introduction}
Among globular proteins,
the equilibrium properties (structure factors, thermodynamic
properties and phase behavior) of aqueus solutions
of lysozyme and $\gamma$-crystalline are the ones, perhaps, most thoroughly
studied.
Numerous recent studies have been reported for these systems from both theoretical
and experimental perspective
(see
\cite{piazza1,piazza2,gogelein,abramo1,abramo2,mezzenga} and
references therein). Much effort, in particular, has
focused on the investigation of the phase behavior of the lysozyme
solution. Experimentally, the liquid--gas phase coexistence for this system was
described by Ishimoto and Tanaka \cite{Ishimoto}.
According to the later studies of Broide et al.  \cite{broide},
this phase coexistence is unstable with respect
to crystallization. Worth mentioning is an important
contribution due to George and Wilson \cite{george}, who
discovered the existence of the `crystallization slot' for the
values of the second osmotic virial coefficient in the vicinity of
the liquid--gas critical point, where one might expect
crystallization of the proteins.

Since globular proteins can be viewed as
colloidal macroions, most of the theoretical studies of protein phase equilibrium
are based on the concepts borrowed from the physics of colloids.
As far as the interaction between two protein macromolecules
is very complicated and, to a large degree, the known theoretical
studies are based on the coarse-grained potential models
\cite{tozzini}. The simplest version of the model,
the so-called one-component model, represents the effect of the solvent and
electrolyte produced by the continuum approximation.
Usually, the corresponding effective interaction is represented by the
hard-sphere (or soft-sphere) interaction combined with long-range
repulsive screened Coulomb interaction and short-range attractive
van der Waals interaction. This is the model utilized in the
Derjaguin-Landau-Verwey-Overbeek (DLVO) theory of the colloidal
stability \cite{DLVO}.
{  Recent computer simulation studies
\cite{abramo1,abramo2,pellicane1,pellicane2} demonstrate that
DLVO model, being not very accurate in reproducing the
short-range behavior of the experimental structure factors at higher
pH values, appears to be quite
successful in describing the
phase behavior of the lysozyme solution.} In particular, DLVO
model was capable of reproducing the flat portion of the
experimental phase diagram in the vicinity of the critical point.
Further progress in the coarse-grained modelling of protein
solutions is related to the introduction of the orientationally
dependent short-range attraction between colloidal particles
\cite{lomakin,linze1,linze2}. The models of this type are aimed at
a more detailed description of the protein molecules taking into
account the existence of charged groups on their surfaces. Phase
behavior and structural properties of these models have been studied
in references
\cite{gogelein,abramo1,abramo2,sear,kalyuzhnyi1,allahyarov,kern,lund,liu,rosch,kalyuzhnyi2}.
Further steps in the detalization of the interaction in protein solutions
are connected with the substitution of the one-component model with the
one that contains multiple components and
takes into account simple salt
ions and solvent molecules. The
presence of a simple electrolyte can be addressed within the framework of
a highly asymmetric electrolyte solution model, which has been extensively
studied using integral equation methods (see \cite{vlachy1} and
references therein) and by explicitly taking into account the effects
of association \cite{k3,k3a,b,k4,k45,k45a,k5}. Finally, the effects due to
solvent molecules can be considered by extending the scheme
developed for simple electrolytes and polyelectrolytes
\cite{vlachy2,vlachy3,k6,k7,k8,k9,k10,k11}

Despite a good performance of the DLVO model in predicting the phase behavior of the
electrolyte solution of lysozyme \cite{abramo1,abramo2}, the authors have not
proposed any theoretical description for its equilibrium properties.
The goal of the present study is to fill this gap and develop simple theories
capable of describing the structural properties and phase behavior of the model,
at least at a qualitative level.
With this goal in mind, we apply several simple liquid state theories,
including the mean spherical approximation (MSA), second order
version of the perturbation theory (TPT), for associating fluids in combination with BH
perturbation theory, and critically assess their performance.
The paper is
organized as follows. In the section~\ref{sec:2} we formulate a potential model
while in section~\ref{sec:3} we discuss the details of the MSA, BH
perturbation theory and TPT. Our results are presented in section~\ref{sec:4} and our
conclusions are given in section~\ref{sec:5}.

\section{The model}
\label{sec:2}

DLVO model
{utilized}
in~\cite{abramo1,abramo2,DLVO}  treats
lysozyme solution as an effective one-component fluid of spherical
particles with the number density $\rho$ interacting via a pairwise
additive potential which is only a function of the interparticle distance $r$.
 Pair interaction between particles consist of (i)
a short-range attractive part of van der Waals term,
\begin{equation}
V_{\rm vdW}(r) =
 - \frac{A_\textrm{H}}{12} \left( \frac{\sigma^2_{\rm dlvo}}{r^2}+\frac{\sigma^2_{\rm DLVO}}{r^2-\sigma^2_{\rm DLVO}}+2 \ln
 \frac{r^2-\sigma^2_{\rm DLVO}}{r^2} \right),
\label{vdW}
\end{equation}
where $\sigma_{\rm DLVO} $ is an effective hard-core diameter and
$A_\textrm{H}$ is the  Hamaker constant, and (ii) a Debye-H\"{u}ckel (DH) term
\begin{equation}
V_{\rm DH}(r) =
  \frac{1}{4 \pi \epsilon_0 \epsilon_r} \left( \frac{Ze}{1+\kappa \sigma_{\rm DLVO}}\right)^2 \frac{\exp\left[-\kappa \left(r-\sigma_{\rm DLVO}\right)\right]}{r},
\label{DH}
\end{equation}
where $Ze$ is a net charge of the lysozyme macromolecule in
electronic units, $\epsilon_0$ is permittivity of vacuum,
$\epsilon_r=86.765-0.3232 \,{\times}\, T(^\circ {\rm C})$ \cite{abramo2}.
The Debye-H\"{u}ckel screening length $\kappa$ is
defined by the expression
\begin{equation}
\kappa =
  \sqrt{\frac{2 I e^2}{\epsilon_0 \epsilon_r k_\textrm{B} T}}\,,
\label{kappa}
\end{equation}
where $I$ is the ionic strength of the solution, which takes into account
the presence of ions due to the buffer and the added salts.
In the present study, we neglect a weak dependence of DH potential on the temperature
and assume for $\kappa$ and $\epsilon_r$ the values calculated at ambient conditions.

The total DLVO potential is written as follows:
\begin{eqnarray}
V_{\rm DLVO}(r)=\left\{
  \begin{array}{ll}
   \infty, & \hbox{$r < \sigma_{\rm DLVO}+\delta$}, \\
    V_{\rm vdW}(r) + V_{\rm DH}(r), \qquad  & \hbox{$r \geqslant  \sigma_{\rm DLVO}+\delta$},
  \end{array}
\right.
\label{Dlvo}
\end{eqnarray}
where the cut-off value $\delta$ is introduced to avoid a singularity
of the van der Waals contribution at $r=\sigma_{\rm DLVO} $ and
corresponds to the thickness of the Stern layer.

\section{Theory}
\label{sec:3}

The properties of the model at hand are studied using MSA,
BH perturbation theory and TPT for associating
fluids. {MSA is a simple analytical approach, which is known for being capable of providing
both structural and thermodynamic properties of a large number of model systems with
sufficient accuracy. Relatively simple and possibly accurate description of thermodynamic
properties of the model can be also achieved within the framework of the BH perturbation theory. Finally,
to account for strong and short-range attraction between particles, which characterizes DLVO potential
model (\ref{Dlvo}), we apply an appropriate combination of the BH perturbation theory and TPT for
associating fluids. The accuracy of each of these approaches is evaluated via comparison of the
theoretical predictions with the corresponding computer simulation and experimental predictions.}

\subsection{MSA}

MSA consists of the Ornstein-Zernike (OZ) equation
\be
{\hat h}(k)={\hat c}(k)+\rho{\hat c}(k){\hat h}(k),
\label{OZ}
\ee
where ${\hat c}(k)$ and ${\hat h}(k)$ are Fourier transforms of the direct and total correlation
functions $c(r)$ and $h(r)$, respectively, and MSA closure relation

\begin{eqnarray}
\left\{
  \begin{array}{ll} c(r)=-\beta \left[V_{\rm vdW}(r)+V_{\rm DH}(r)\right],
   \hspace{10mm}  & \hbox{$r \geqslant \sigma_{\rm DLVO}+\delta$}, \\
     h(r)=-1, \qquad
      \,\, & \hbox{$r <  \sigma_{\rm DLVO}+\delta$.}
  \end{array}
\right.
\label{MSA}
\end{eqnarray}

For the sake of analytical description, we approximate the vdW part of the potential $V_\textrm{vdW}(r)$
using one-Yukawa potential, i.e.,
\begin{equation}
V_{\rm vdW}(r) \equiv V_{\rm Y}(r)=
  - A_\textrm{Y} \frac{\exp[-\kappa_\textrm{Y} (r-(\sigma_{\rm DLVO}+\delta)]}{r} \,.
\label{Y}
\end{equation}

As a result, the pair potential outside the hard core is represented by
the two-Yukawa potential and we
{have used an analytical solution of the MSA due to Blum and Hoye}
\cite{BlumHoye}. The structural and thermodynamic properties have been calculated
utilizing this solution and using a closed form of analytical expressions presented in \cite{arrieta}.

\subsection{Barker-Henderson perturbation theory}

Here, we utilize the second-order BH perturbation theory. Within the framework of
BH perturbation theory \cite{BH}, Helmholtz free energy $F$ of the system per particle is given by
the following expression
\begin{equation}
\frac{F}{Nk_\textrm{B}T} = \frac{F^\textrm{ideal}}{Nk_\textrm{B}T}+\frac{F^\textrm{HS}}{Nk_\textrm{B}T}
+\frac{F_1}{Nk_\textrm{B}T}+\frac{F_2}{Nk_\textrm{B}T}\,,
\label{BH2}
\end{equation}
where $k_\textrm{B}$ is the Boltzmann constant, $F^\textrm{ideal}$ is the ideal gas free energy
\begin{equation}
\frac{F^\textrm{ideal}}{Nk_\textrm{B}T} = \ln{\left( \Lambda\rho\right)} -1 \,,
\label{BH21}
\end{equation}
$\Lambda$ is the thermal de Broglie wavelength,
$F^\textrm{HS}$ is the hard-sphere Helmholtz free energy and $F_1$ and $F_2$ are the first- and
second-order contributions to Helmholtz free energy of the system. Here, for $F^\textrm{HS}$ we have
used the Carnahan-Starling extension
\begin{equation}
\frac{F^\textrm{HS}}{Nk_\textrm{B}T} = \frac{4\eta-3\eta^2}{(1-\eta)^2}
 \,,
\label{BH22}
\end{equation}
where $\eta=\frac{\pi}{6}\rho\sigma_0^3$, $\sigma_0=\sigma_{\rm DLVO}+\delta$ and for
$F_1$ and $F_2$, we have
\begin{align}
\label{BH23}
\frac{F_1}{Nk_\textrm{B}T} &= \frac{2\pi\rho}{k_\textrm{B}T}\int_{\sigma_0}^{\infty}
V_{\rm DLVO}(r) g^\textrm{HS}(r)r^2 \rd r \,, \\
%
\frac{F_2}{Nk_\textrm{B}T} &=-{\frac{\pi\rho}{\left(k_\textrm{B}T\right)^2}}
\kappa^\textrm{HS}
\int_{\sigma_{0}}^{\infty} V_{\rm DLVO}^2(r) g^\textrm{HS}(r)r^2 \rd r \,,
\label{BH33}
\end{align}
where $\kappa^\textrm{HS}$ is the compressibility of the hard-sphere fluid. Here, we use the
corresponding Carnahan-Starling expression \cite{CS}.

Using the above expression (\ref{BH2}) for Helmholtz free energy of the system, all the rest
thermodynamical properties (pressure and chemical potential) can be derived using the standard
thermodynamical relations. As for any perturbation theory, the BH model works best if the attractive
potential is not large in magnitude. For potentials with strong attractions, other approaches
are more suitable.

\subsection{Thermodynamic perturbation theory for associating fluid}

To account for the strong attraction seen in the DLVO potential of lysozyme,
we combine the BH perturbation
theory and thermodynamic perturbation theory (TPT) for associating fluids
{with spherically symmetric interaction}
\cite{Kalyuzhnyi2007}.
Following the earlier studies \cite{k3,k4,Kalyuzhnyi2-2}, the total pair
potential of the system (\ref{Dlvo}) is represented as a sum of the reference and associating
pieces, i.e.,
\be
V_{\rm DLVO}(r)=V_{\rm DLVO}^{(\textrm{ref})}(r)+V_{\rm DLVO}^{(\textrm{as})}(r)\,,
\label{split0}
\ee
where
$V_{\rm DLVO}^{(\textrm{ref})}(r)=V_{\rm DLVO}(r)-V_{\rm DLVO}^{(\textrm{as})}(r)$ and
\begin{eqnarray}
V_{\rm DLVO}^{(\textrm{as})}(r)=\left\{
  \begin{array}{ll}
   V_{\rm DLVO}(r)-V_0^{(\textrm{as})},
\hspace{5mm}  & \text{for} \quad V_{\rm DLVO}(r)\leqslant V_0^{(\textrm{as})}, \\
    0, \hspace{10mm} & \text{otherwise}.
  \end{array}
\right.
\label{splitting}
\end{eqnarray}
Here, $V_{\rm DLVO}(\sigma_0) < V_0^{(\textrm{as})} < 0$,
and we assume that $V_0^{(\textrm{as})}$ is temperature dependent.
A particular choice for the potential splitting
parameter $V_0^{(\textrm{as})}$ is discussed below. According to the splitting of the total DLVO
potential (\ref{split0}), the system Helmholtz free energy within the framework of the TPT
\cite{Kalyuzhnyi2007} is as follows:
\be
F=F^\textrm{ref}+F^\textrm{as},
\label{ATPT}
\ee
where for the free energy of the reference system $F^\textrm{ref}$ we have used the second-order expression
(\ref{BH2}) with DLVO pair potential $V_{\rm DLVO}(r)$ substituted by the reference potential
$V_{\rm DLVO}^{(\textrm{ref})}(r)$, and thus for the associative part $F^{(\textrm{as})}$ we have
\be
{\frac{F^\textrm{as}}{Vk_\textrm{B}T}}=\rho\ln{\left({\frac{\chi_0}{\rho}}\right)}
+{\frac{1}{2}}\chi_{m-1}
{\frac{\chi_1-\chi_0}{\chi_0}},
\label{TPT1}
\ee
where
\be
\chi_l=\chi_0\sum_{k=0}^l{\frac{1}{k!}}\left({\frac{\chi_1-\chi_0}{\chi_0}}\right)^k,
\qquad \text{for} \qquad l=2,\ldots,m,
\label{TPT2}
\ee
$\chi_0$ and $\chi_1$ satisfy the following set of equations
\begin{eqnarray}
\left\{
  \begin{aligned}
{\frac{\chi_1-\chi_0}{\chi_0}}&=\left[\rho-{\frac{1}{ m!}}{\frac{\left(\chi_1-\chi_0\right)^m}{\chi_0^{m-1}}}\right]K,\\
    \rho\chi_0^{m-1}&=\sum_{k=0}^m{\frac{\chi_0^k}{ \left(m-k\right)!}}\left(\chi_1-\chi_0\right)^{m-k},\\
  \end{aligned}
\right.
\label{setg}
\end{eqnarray}
where
\be
K=4\pi g^{(\textrm{ref})}(\sigma_0)\int\left[\exp{\left(-{\frac{V_{\rm DLVO}^\textrm{as}(r)}{ k_\textrm{B}T}}\right)}-1\right]r^2\, \rd r,
\label{K}
\ee
$m$ is a maximum number of bonds per particle allowed
and $g^{(\textrm{ref})}(\sigma_0)$ is the contact value of the radial distribution function of the reference
system. The latter quantity was obtained using the parametrization of the contact value of the radial
distribution function for the hard-sphere square-well fluid \cite{SW}.

The knowledge of the free energy of a system (\ref{ATPT}) enables one to calculate thermodynamical
properties of interest using standard thermodynamical relations.
The version of the TPT approach discussed above has two input parameters, i.e., the maximum number
of bonds $m$ and the potential splitting parameter $V_0^{(\textrm{as})}$. In general, the number of nearest neighbours
for the model at hand could be up to 12, thus one can assume $m=12$. In this case, the value of
$V_0^{(\textrm{as})}$, which ensures the saturation of the associative potential $V_{\rm DLVO}(r)$ for $m=12$, can
be chosen to be relatively large in comparison
{ with the contact value of the DLVO potential}
$V_{\rm DLVO}(\sigma_0)$. However, according to the
earlier studies~\cite{wtpt}, `single bond' approximation utilized here is accurate only
for relatively small values of $m$. In addition, the probability of bonding of 12 particles
simultaneously at the densities where the liquid--gas separation occurs is small; thus, the
optimal choice for $m$ and $V_0^{(\textrm{as})}$ requires a certain compromise. In the present study we assume
that $m=3$ and
{  the reduced value of the associative potential at the contact
$V_{\rm DLVO}^{*(\textrm{as})}(\sigma_0)=V_{\rm DLVO}^{(\textrm{as})}(\sigma_0)/k_\textrm{B}T$
is constant. In addition, to provide an accurate description of the reference system at the critical
point, we also assume that $V_{0}^{(\textrm{as})}=k_\textrm{B}T_\textrm{cr}$, where $T_\textrm{cr}$ is the critical temperature. Combining
the latter two assumptions, we have}
\be
{\frac{V_0^{(\textrm{as})}}{ k_\textrm{B}T}}=V_{\rm DLVO}(\sigma_0)\left({\frac{1}{ k_\textrm{B}T}}-{\frac{1}{ k_\textrm{B}T_\textrm{cr}}}\right)-1.
\label{V0}
\ee
For $T=T_\textrm{cr}$ we have $V_0^{(\textrm{as})}=-k_\textrm{B}T_\textrm{cr}$, i.e., for the critical temperature, the minimum value of
the reference system potential $V_{\rm DLVO}^{(\textrm{ref})}(r)$ is equal to $-k_\textrm{B}T_\textrm{cr}$.
{For this value of the potential minimum, the BH approach is expected to be sufficiently
accurate}.
Although our choice for $m$ and $V_0^{(\textrm{as})}$ is rather empirical, the theory proposed is self-contained because there is no need in the input from outside.
{In particular, critical temperature $T_\textrm{cr}$ (and critical density $\rho_\textrm{cr}$), which enter the expression
for $V_0^{(\textrm{as})}$ (\ref{V0}), is obtained as usual from the
solution of the set of two equations, which requires the first and the second derivatives of the
pressure with respect to the density to be equal to zero, i.e.,
\begin{eqnarray}
\left\{
  \begin{aligned}
{\frac{\partial P(\rho,T_\textrm{cr})}{\partial\rho}}\bigg|_{\rho=\rho_\textrm{cr}}&=0,\\
{\frac{\partial^2 P(\rho,T_\textrm{cr})}{\partial\rho^2}}\bigg|_{\rho=\rho_\textrm{cr}}&=0.\\
\end{aligned}
\right.
\end{eqnarray}
}

\section{Results and discussion}
\label{sec:4}

In this section we present our numerical results for the structural properties and phase behavior
of lysozyme and $\gamma$-crystalline aqueous electrolyte solutions. In both cases, we use the DLVO
model with $\sigma_{\rm DLVO}=37.08$~{\AA} for lysozyme and $\sigma_{\rm DLVO}=37.8$~{\AA} for
$\gamma$-crystalline. For the Stern-layer thickness and Hamaker constant, the following values
have been used \cite{abramo2}: $\delta=1.8$~{\AA} and $A_\textrm{H}=18.8$~kJ/mol.

\begin{figure}[!t]
\centerline{
\includegraphics[width=0.49\textwidth]{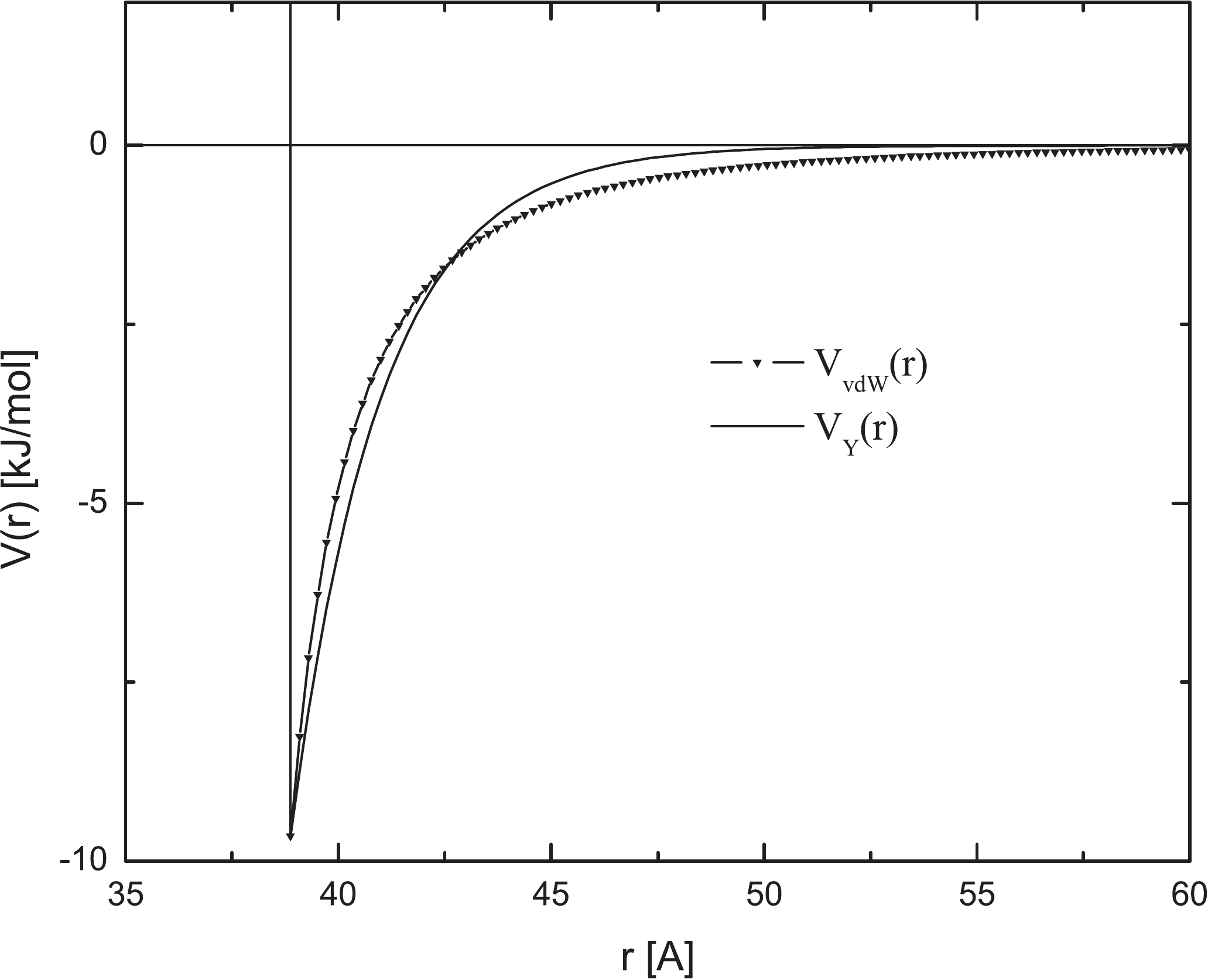}
}
\caption{Van der Waals contribution
to the DLVO potential (line with triangles) [see (\ref{vdW})] and Yukawa
substitution (full line) of van der Waals term.}
\label{fig1}
\end{figure}
\begin{figure}[!b]
\centerline{
\includegraphics[width=0.49\textwidth]{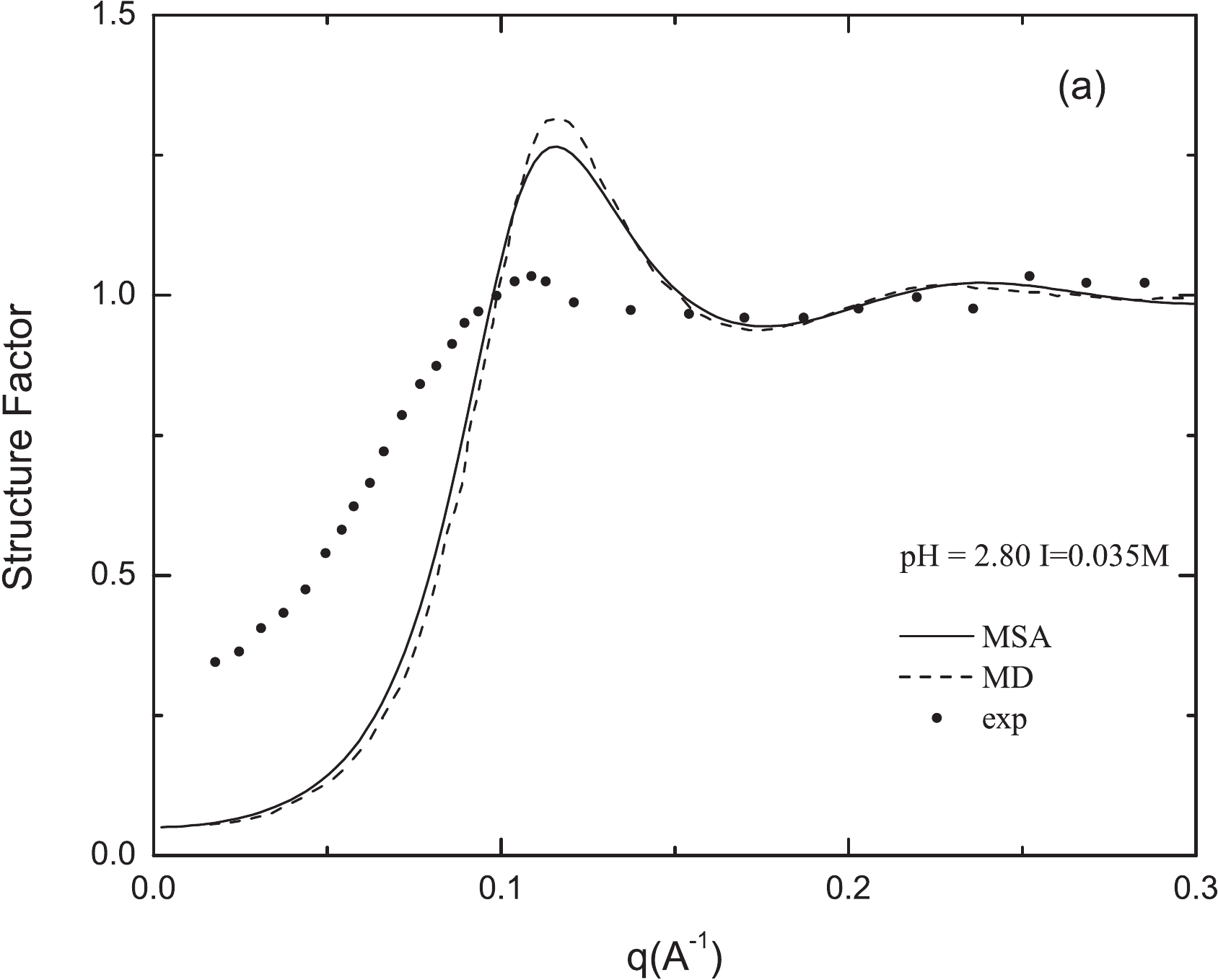}
\includegraphics[width=0.49\textwidth]{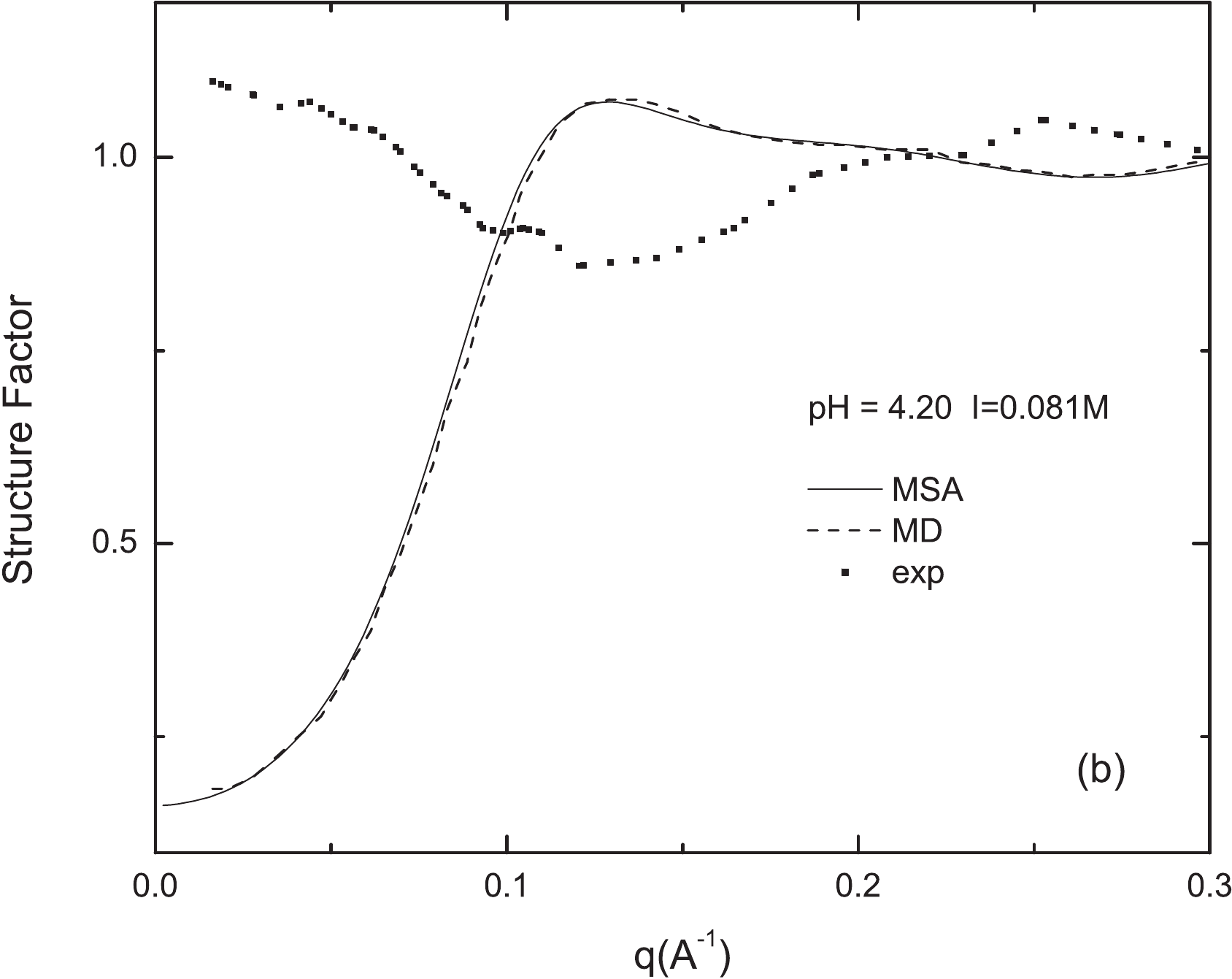}
}
\vspace{2mm}
\centerline{
\includegraphics[width=0.49\textwidth]{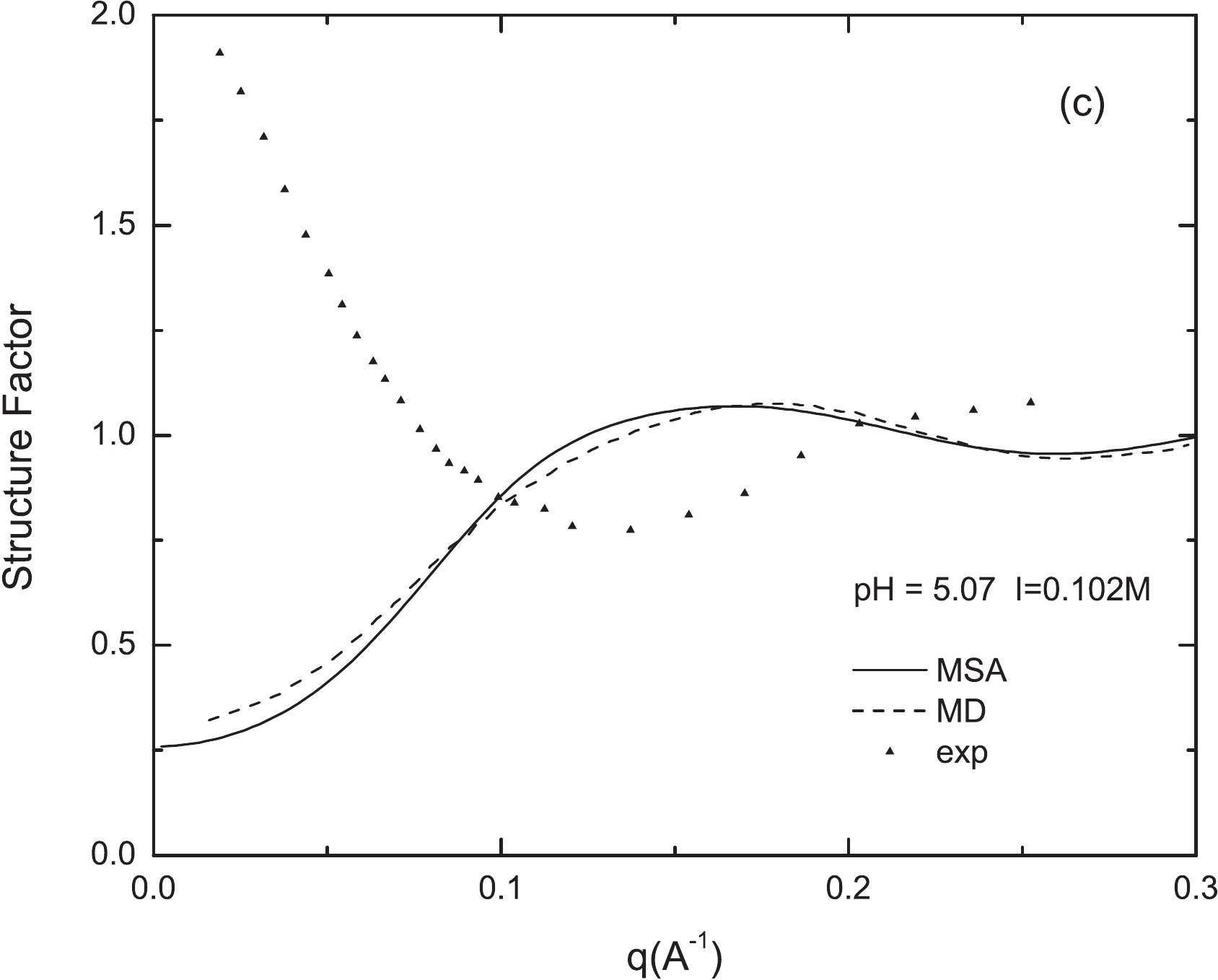}
}
 \caption{Structure factor of the
lysozyme aqueous electrolyte solution at different pH.} \label{fig2}
\end{figure}

\begin{figure}[!t]
\centerline{
\includegraphics[width=0.61\textwidth]{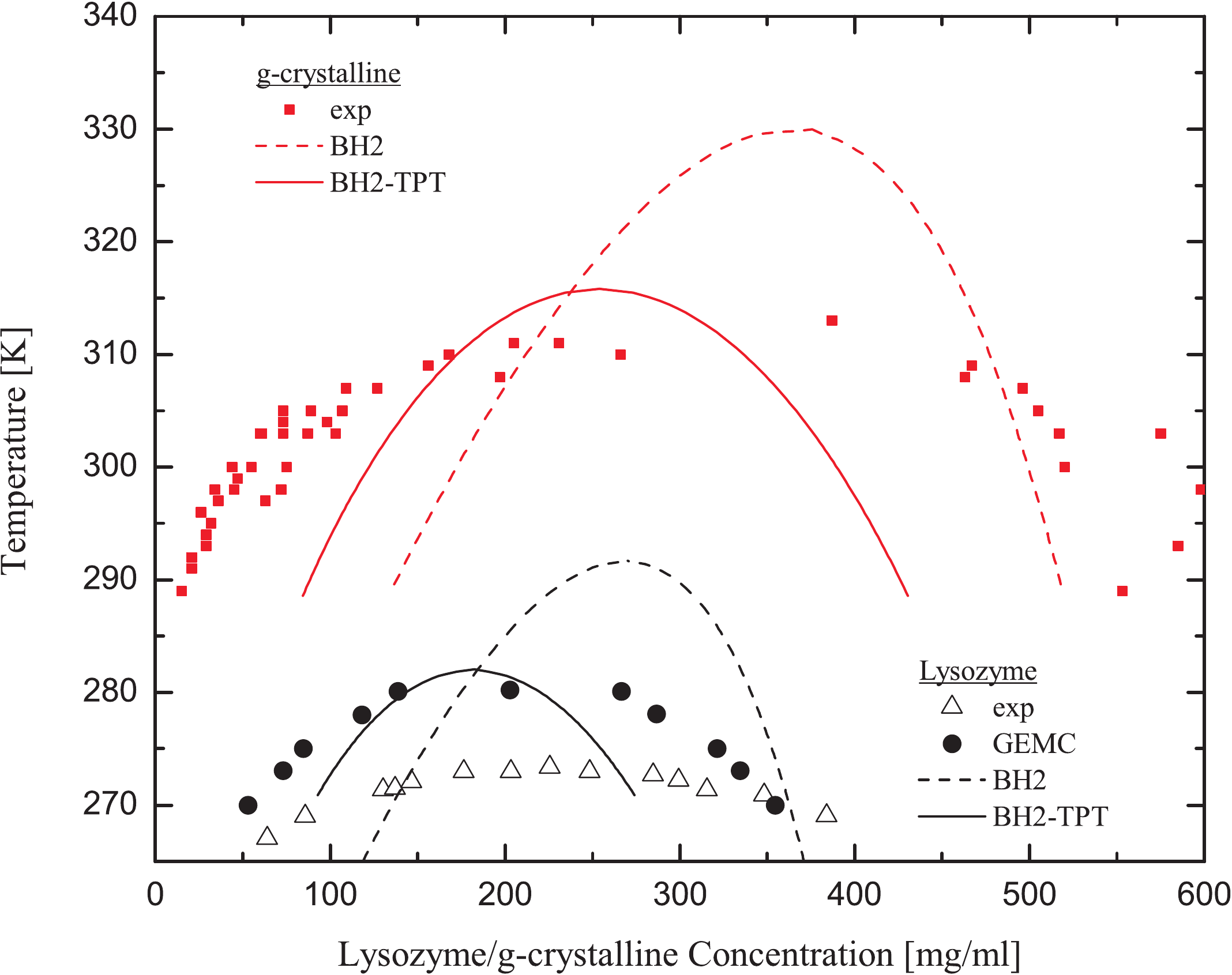}
}
 \caption{(Color online) Liquid--gas phase diagrams of the
lysozyme aqueous electrolyte solution at $\text{pH}=6.0$ and ionic strength $I=0.6$~M (black color) and $\gamma$-crystalline aqueous electrolyte solution
at $\text{pH}=7.1$ and  $I=0.24$~M (red color).
{  Experimental values for the critical concentration and temperature for lysozyme
\cite{exp1} and $\gamma$-crystalline solutions \cite{exp2} are ($230\pm 10$~mg/ml, 273~K)
and ($289\pm 20$~mg/ml, 311~K), respectively.}}
 \label{fig3}
\end{figure}

The structural properties of the lysozyme solution [static structure factor $\,S(k)\,$ and
radial distribution function $g(r)$] were calculated using MSA (\ref{MSA}). Parameters for the
Yukawa potential (\ref{Y}) representing $V_{\rm vdW}(r)$ were chosen using `best eye fit' method
supplemented by the equality of both potentials at the contact distance, i.e.,
$V_\textrm{Y}(\sigma_0)=V_{\rm vdW}(\sigma_0)$. In figure~\ref{fig1} we compare $V_\textrm{Y}(r)$ and $V_{\rm vdW}(r)$ with
the following choice of the Yukawa potential parameters: $A_\textrm{Y}=375.19$ kJ/mol$\cdot${\AA} and
$\kappa_\textrm{Y} = 0.45$~{\AA}$^{-1}$.
{In figure~\ref{fig2} we
compare the corresponding theoretical and computer simulation \cite{abramo2}
results for the structure factor $S(k)$
at the values of pH used in the {corresponding experimental study},
i.e., $\text{pH}=2.8$, 4.2, 5.07.
These values of pH correspond to the following three values of the ionic strength
(see table~1 of reference \cite{abramo2}), i.e., $I=0.035$~M, 0.081~M, 0.102~M, respectively.}
In addition, we also show the
corresponding experimental results \cite{abramo2} for $S(k)$. Very good agreement is
observed between
theoretical and computer simulation results  for the structure factor at all values
of pH and $I$ studied.
{Although DLVO model is not very accurate in describing the short-range behavior of the structure
factor at higher values of pH (figure~\ref{fig2}), its predictions for the phase behavior are in reasonable
agreement with experimental predictions (see figure~\ref{fig3}).}
These features of the DLVO model (\ref{Dlvo}) are known and have been discussed earlier
\cite{abramo2}.
In figure~\ref{fig3} we show our theoretical results for the liquid--gas phase diagram of the
lysozyme solution at $\text{pH}=6.0$ and ionic strength $I=0.6$~M (black color) and $\gamma$-crystalline aqueous electrolyte solution
at $\text{pH}=7.1$ (which correspond to the protein
overall charge $Z=1$) and ionic strength $I=0.24$~M (red color). These results were obtained using
second-order BH theory
(\ref{BH2}) and our version of the TPT (\ref{ATPT}). Unfortunately, for the employed parameters of
 the DLVO model, the
 MSA does not have a convergent solution in the range of temperatures
and densities, where one would expect the location of binodals. In addition,
in the same figure we show computer simulation results and the results of the experiment \cite{exp1, exp2}.
Here, computer simulation predictions are much more accurate in comparison with the predictions for
the structure factors demonstrated in the previous figure. Our results obtained using the BH
perturbation theory, are much less accurate, giving the values for both critical
temperature and critical density that are too large. In addition, BH phase diagram is too narrow
in comparison with
the computer simulation and experimental phase diagrams. Predictions of our combined BH and
thermodynamic perturbation theory are in reasonable agreement with the computer simulation data.
{  Good agreement is observed for
 the critical temperature. Slightly less accurate are the predictions for the critical concentration}.
The overall shape of the
phase diagram is still too narrow, although it is now much closer to the shape of computer
simulation phase diagram.

\section{Conclusions}
\label{sec:5}

In this article we studied the structural properties and phase behaviour of the DLVO model of
lyso\-zyme and $\gamma$-crystalline aqueous electrolyte solutions using MSA, second-order BH
perturbation theory and a combined approach based on the BH theory and TPT for associating fluids.
Theoretical results are compared with computer simulations and experimental results.
Predictions of the MSA for the structure factor of lysozyme solution are in good agreement with the
corresponding computer simulation predictions.
{  However, MSA does not have a convergent solution in the range of the temperatures
and densities, where one would expect the location of the corresponding MSA binodals. We conclude
that MSA is inappropriate for the phase behavior of the DLVO-type models of lyzosyme and, perhaps,
other globular proteins.
}
Among the theories used to describe the phase behaviour of the lysozyme and
$\gamma$-crystalline solutions, only a combined BH-TPT approach provides a reasonable
  qualitative agreement with computer simulations and experimental description.
We expect that a further improvement of the theory can be achieved using association
concepts in combination with a more detailed description of the protein molecules taking into account
the existence of the charged groups on their surfaces. Corresponding studies are underway and
results will be reported in due course \cite{privat}.

\section*{Acknowledgement}
Author gratefully acknowledge Professor Yu.~Kalyuzhnyi for discussions that led to this work and for providing a source code of the multi-Yukawa MSA code.



\ukrainianpart

\title{Теоретичне дослідження фазової поведінки моделі ДЛВО водних розчинів електролітів лізоциму і гамма-кристаліну}
\author{Р. Мельник}
\address{
Інститут фізики конденсованих систем Національної академії наук України, \\
вул. І.~Свєнціцького, 1, 79011 Львів, Україна
}
%
%
%

\makeukrtitle

\begin{abstract}
\tolerance=3000%
Середньосферичне наближення (ССН), теорія збурень другого порядку Баркера-Гендерсона (БГ) і термодинамічна теорія збурень (ТТЗ) для асоціативних рідин в комбінації з теорією збурень БГ застосовані до вивчення структурних властивостей та фазової поведінки моделі Дєрягіна-Ландау-Вервея-Овербека (ДЛВО) водних розчинів електролітів лізоциму і гамма-кристаліну. Результати ССН для структурних факторів добре узгоджуються з відповідними комп'ютерними розрахунками. Узгодження між теоретичними результатами для фазової діаграми рідина--газ і відповідними експериментами та комп'ютерними симуляціями є менш задовільним. Кращого узгодження для фазових діаграм дозволяє досягнути комбінований БГ-ТТЗ підхід.
\keywords модель ДЛВО, потенціал Юкави, водні розчини лізоциму і гамма-кристаліну, середньосферичне наближення

\end{abstract}

\end{document}